\documentclass[12pt,preprint]{aastex}

\usepackage{psfig}
\usepackage{epsfig}
\usepackage{amsmath}

\begin{document}

\title{{\em Spitzer} and {\em Hubble} Constraints on the Physical
Properties of the z$\sim$7 Galaxy Strongly Lensed by
Abell~2218\altaffilmark{1,2}}

\author{
E.~Egami\altaffilmark{3},
J.-P.~Kneib\altaffilmark{4,5},
G.~H.~Rieke\altaffilmark{3},
R.~S.~Ellis\altaffilmark{5},
J.~Richard\altaffilmark{4,5},
J.~Rigby\altaffilmark{3},
C.~Papovich\altaffilmark{3},
D.~Stark\altaffilmark{5},
M.~R.~Santos\altaffilmark{6},
J.-S.~Huang\altaffilmark{7},
H.~Dole\altaffilmark{3},
E.~Le~Floc'h\altaffilmark{3},
and P.~G.~P\'{e}rez-Gonz\'{a}lez\altaffilmark{3}
}

\altaffiltext{1}{This work is based in part on observations made with
the Spitzer Observatory, which is operated by the Jet Propulsion
Laboratory, California Institute of Technology under NASA contract
1407. Support for this work was provided by NASA through Contract
Number 960785 issued by JPL/Caltech.}
\altaffiltext{2}{Based on observations made with the NASA/ESA Hubble
  Space Telescope, obtained at the Space Telescope Science Institute,
  which is operated by the Association of Universities for Research in
  Astronomy, Inc., under NASA contract NAS 5-26555.  These
  observations are associated with program \#9452.}
\altaffiltext{3}{Steward Observatory, University of Arizona, 933
  N. Cherry Avenue, Tucson, AZ85721; (eegami, grieke, jrigby,
  papovich, hdole, elefloch, pgperez)@as.arizona.edu}
\altaffiltext{4}{Observatoire Midi-Pyr\'en\'ees, UMR5572,
     14 Avenue Edouard Belin, 31000 Toulouse, France; jrichard@ast.obs-mip.fr}
\altaffiltext{5}{Department of Astronomy, California Institute of
  Technology, 105-24, Pasadena, CA91125; (kneib, rse, dps)@astro.caltech.edu}
\altaffiltext{6}{Institute of Astronomy, Madingley Road, Cambridge,
CB3 0HA, UK; mrs@ast.cam.ac.uk} 
\altaffiltext{7}{Harvard-Smithsonian Center for
Astrophysics, 60 Garden Street, Cambridge, MA02138; jhuang@cfa.harvard.edu}

\begin{abstract}

We report the detection of a $z \sim 7$ galaxy strongly lensed by the
massive galaxy cluster Abell 2218 ($z=0.175$) at 3.6 and 4.5 $\mu$m
using the {\em Spitzer Observatory} and at 1.1 $\mu$m using the {\em
Hubble Space Telescope}.  The new data indicate a refined photometric
redshift in the range of $6.6-6.8$ depending on the presence of
Ly$\alpha$ emission.  The spectral energy distribution is consistent
with having a significant Balmer break, suggesting that the galaxy is
in the poststarburst stage with an age of at least $\sim$ 50 Myr and
quite possibly a few hundred Myr. This suggests the possibility that
a mature stellar population is already in place at such a high
redshift.  Compared with typical Lyman break galaxies at $z \sim 3-4$,
the stellar mass is an order of magnitude smaller ($\sim 10^{9}
M_{\sun}$), but the specific star formation rate (star formation
rate/$M_{star}$) is similarly large ($> 10^{-9}$ yr$^{-1}$),
indicating equally vigorous star-forming activity.

\end{abstract}


\section{Introduction}

Locating and characterizing the first sub-galactic sources which may
have been responsible for completing cosmic reionization and ending
the ``Dark Ages" represents the latest frontier in observational
cosmology.  Although gravitational instability theory allows the early
formation of massive halos \citep[e.g.,][]{Barkana00}, the
complexities of gas cooling, star formation, feedback and the
clumpiness of the intergalactic medium make observational predictions
highly uncertain \citep{Stiavelli04}. Data on representative early
sources are needed to better understand when reionization occurred as
well as what the luminosity and mass functions are of the sources
responsible for the reionization.

Searching for representative sources at redshift ($z$) of 6--10, the
range considered to be the final stages of the reionization era, is a
major observational challenge, not only because of the large
luminosity distances involved but also since (as predicted by the
hierarchical model of galaxy formation) the most likely systems to be
found during this period have lower stellar masses than those at more
moderate redshifts.  The high magnification associated with strong
gravitational lensing by massive foreground clusters offers a unique
opportunity for detecting such galaxies.  In recent years, this
approach has been employed specifically to find galaxies during the
reionization era, and has produced a number of successes
\citep{Ellis01,Hu02,Santos04,Kneib04a}. The detection of a $z \sim 10$
galaxy by \citet{Pello04}, however, is controversial
\citep{Weatherley04,Bremer04}.

In this Letter, we return to analyze in more detail the physical
characteristics of the triply lensed $z \sim 7$ source in Abell~2218
($z=0.175$) reported by \citet{Kneib04a}.  The original redshift
estimate was based on three pieces of observational evidence obtained
for the two brighter components $a$ and $b$: (1) reflection symmetry
with respect to the well-defined $z \ga 6$ critical lines, (2)
photometric redshift suggesting $6.6 < z < 7.1$, and (3) a tentative
spectroscopic detection of a continuum break at 9800 \AA, which, if
due to Gunn-Peterson absorption at Ly$\alpha$, would indicate
$z=7.05$.  Here, we report the {\em Spitzer} detection at 3.6 and 4.5
$\mu$m of these two components as well as the new 1.1 $\mu$m detection
by the {\em Hubble Space Telescope} ({\em HST}).  With these new data,
we will refine the photometric redshift and examine the physical
properties of this lensed galaxy.

\section{Observations}

The 3.6 and 4.5 $\mu$m images were obtained on UT 2004 January 2 with
the InfraRed Array Camera (IRAC; \citet{Fazio04}) on the {\em Spitzer
Observatory}.  In each filter, twelve 200-second images were taken
with the small-step cycling dither pattern.  The Basic Calibrated Data
(BCD) images were combined using a custom IDL mosaicking routine with
the final pixel scale of 0\farcs6 pixel$^{-1}$, half of the instrument
pixel size.  The IRAC observations also provided 5.8 and 8.0 $\mu$m
images of the same field simultaneously, but the $z \sim 7$ galaxy was
not detected at these longer wavelengths.

The 1.1 $\mu$m images were obtained on UT 2004 March 23 with the Near
Infrared Camera and Multi-Object Spectrometer (NICMOS) on {\em HST}.
A total of nine 2304 second (using the SPARS256 sequence) and one 1472
second (using the SPARS64 sequence) exposures were taken with the
F110W filter using the NIC3 camera, which has a pixel scale of
0\farcs2 pixel$^{-1}$.  

The central wavelengths and full widths at 20\% of the transmission
peak are listed in Table~\ref{flux} for all the observed bands based
on the total transmission curves.

\section{Results}

\begin{figure}
  \plotone{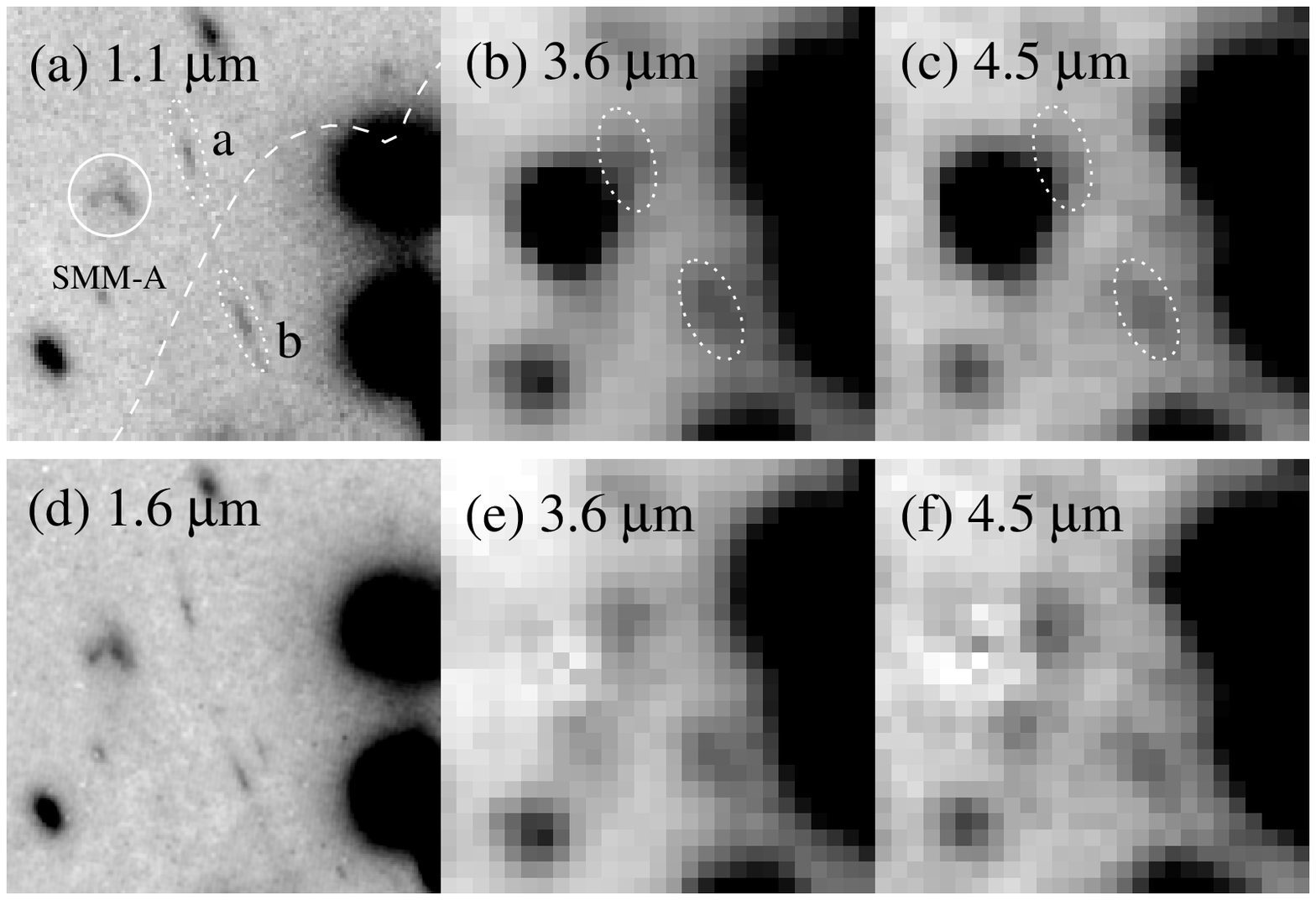}

  \caption{{\em HST}/NICMOS and {\em Spitzer}/IRAC images of the
  $z\sim7$ lensed pair: (a) New NICMOS 1.1 $\mu$m image.  Components
  $a$ and $b$ as well as the z=2.5 submillimeter source SMM-A are
  marked.  The dashed line indicates the $z \ga 6.5$ critical curves;
  (b) IRAC 3.6 $\mu$m image; (c) IRAC 4.5 $\mu$m image; (d) NICMOS 1.6
  $\mu$m image presented in \citet{Kneib04a}; (e) IRAC 3.6 $\mu$m
  image with SMM-A subtracted; (f) IRAC 4.5 $\mu$m image with SMM-A
  subtracted. Each image is 16\arcsec\ on a side.  In the panels (a)
  and (d), north is up and east is to the left.  In the panels (b),
  (c), (e), and (f), we preserved the original array orientation,
  which is 11$\degr$ rotated counter-clockwise from that of the panels
  (a) and (b), in order to display the results of the SMM-A
  subtraction accurately.  The dashed-line ellipses shown in (b) and
  (c) around component $b$ indicate the areas used for photometry.
  Similar ellipses are drawn around component $a$ to mark its
  position. \label{image}}
\end{figure}

Figure~\ref{image} shows the {\em HST}/NICMOS and {\em Spitzer}/IRAC
images.  Components $a$ and $b$ of the galaxy reported by
\citet{Kneib04a} are clearly detected at 1.1 $\mu$m
(Figure~\ref{image}a).  At 3.6 and 4.5 $\mu$m, component $b$ is
clearly detected while component $a$ is seen as a faint extension to
the northwest of the $z=2.5$ lensed submillimeter source SMM-A (SMM
J16359+6612.6) discussed by \citet{Kneib04b} (Figure~\ref{image}b and
c).  For comparison, the {\em HST}/NICMOS 1.6 $\mu$m image by
\citet{Kneib04a} is also displayed (Figure~\ref{image}d), which shows
the two components and SMM-A with a higher signal-to-noise ratio.
When the light from SMM-A is subtracted using a two-dimensional
elliptical Gaussian, component $a$ shows up in the IRAC images
(Figures~\ref{image}e and f).

At 1.1 $\mu$m, photometry was performed in a manner consistent with
that in \citet{Kneib04a}, giving the total fluxes of components $a$
and $b$.  In the IRAC bands, photometry was performed only for
component $b$ because the flux measurements with component $a$ are
highly uncertain due to the subtraction of SMM-A.  For the photometry
of component $b$, we used an elliptical aperture with semi-major and
semi-minor axes of 2\arcsec\ and 1\arcsec, respectively, and aligned
the major axis along the position angle of component $b$ (23\degr\
East of North).  The sky level was measured using similar ellipses to
define the inner and outer boundaries of a sky area with a semi-major
axis of 2 and 3.5 pixels, respectively.  There is a slight sky
background gradient along the east-west direction due to the bright
source to the west, but the component b is far enough from this source
that the gradient is quite linear across component b and the
surrounding sky area.  Therefore, we expect this gradient to be
removed if we derive the local sky level as the average of all the sky
pixel values.  We also visually inspected the image cross sections,
and confirmed that the derived sky level is a good estimate around the
position of component b.  The photometric uncertainty is based on the
scatter of sky pixel values, and our estimates are conservative in
that the scatter not only includes the random sky pixel noise but is
also inflated by the sky gradient along the east-west direction.

Because of the large point spread functions at 3.6 and 4.5 $\mu$m
(FWHM $\sim$ 1\farcs7), the elliptical aperture misses a significant
fraction of the total source flux.  By convolving the F160W image with
the IRAC PSFs, we determined that the flux measured with the
elliptical aperture needs to be multiplied by a factor of 1.4 to
account for this missing flux (i.e., aperture correction).

The measured flux densities at 1.1, 3.6, and 4.5 $\mu$m are listed in
Table~\ref{flux} together with the previous {\em HST} measurements
presented in \citet{Kneib04a}.  The 1.1 and 1.6$\mu$m images show a
faint source north-west of component $b$.  This source is also seen in
the F606W, F814W, and F850LP images \citep{Kneib04a}.  The {\em
Spitzer} images provide enough spatial resolution to conclude that
there is very little flux emitted by this bluer and less distant
source at 3.6 and 4.5 $\mu$m.

\section{Discussion}

The new photometric data permit us to construct the spectral energy
distribution (SED) of component $b$, which we compare with the model
SEDs simulated using the stellar population synthesis code GALAXEV
\citep{Bruzual03}.  A grid of models was calculated with the
parameters described in Table~\ref{model}.  The intrinsic (i.e.,
unmagnified) flux densities were assumed to be 25 times smaller than
the observed values \citep{Kneib04a}.  For each model, the best-fit
SED was sought by varying age and redshift and minimizing $\chi^{2}$.
The predicted fluxes were calculated from the model SEDs with
appropriate total system throughput curves.  The overall normalization
of a model SED was determined by the F160W flux density measurement,
which has the highest signal-to-noise ratio. This leaves five filters
(F814W, F850LP, F110W, IRAC 3.6 and 4.6$\mu$m) to constrain the SED
shape.  Given that little is known about the mode of star formation at
$z > 6$, we tried to bracket various parameter ranges by calculating
models for various star formation histories (see Table~\ref{model}).
The GALAXEV results were cross-checked against Starburst99
\citep{SB99}, which gave similar results.

\begin{figure}
  \epsscale{0.7}
  \plotone{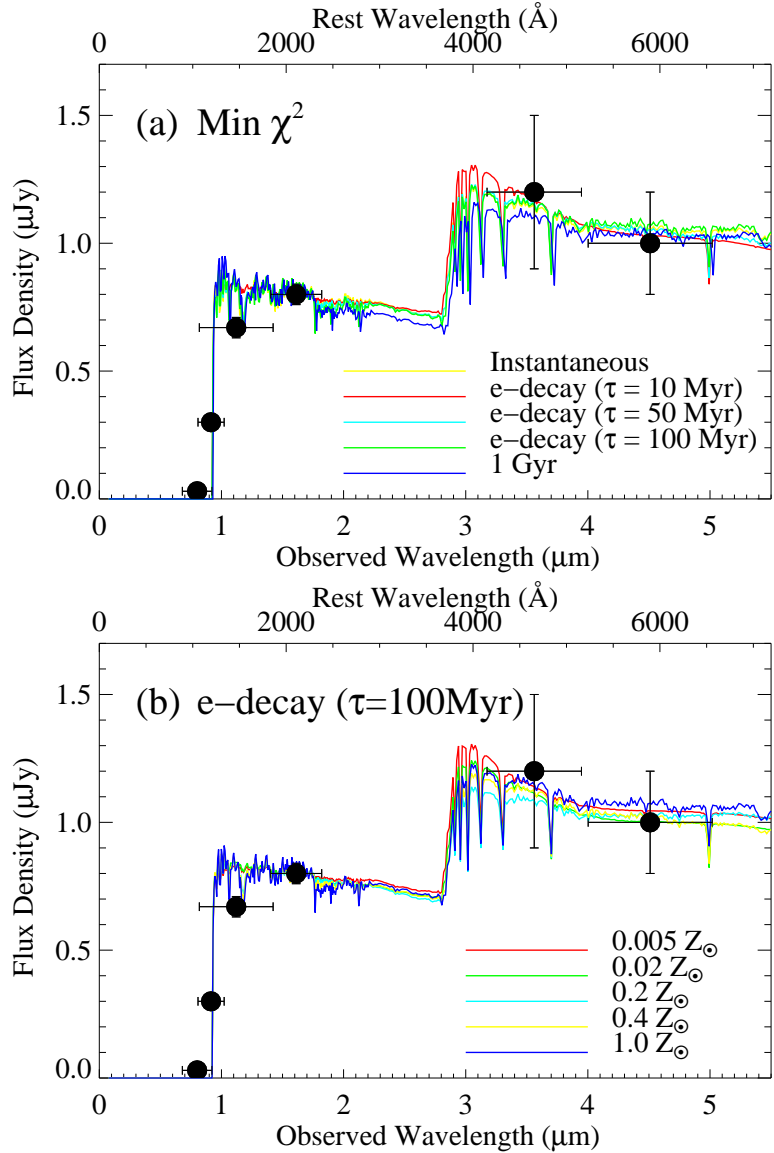}

  \caption{SED model fits to the observed SED of the $z \sim 7$ galaxy
  (component $b$): (a) The best (i.e., minimum $\chi^{2}$) model for
  each star formation history; (b) the exponentially decaying
  starburst model ($\tau = 100$ Myr) for a range of metallicities.
  The restframe wavelength at $z = 6.65$ is also shown. The error bars
  indicate the $\Delta\lambda$ and flux density uncertainty of each
  band shown in Table~\ref{flux}. The measured flux at 1.1 $\mu$m is
  expected to be significantly lower than the true continuum level
  because the F110W filter passband extends below 1216 \AA\ in the
  restframe.  \label{sed}}
\end{figure}

Figure~\ref{sed}a shows the best (i.e., minimum $\chi^{2}$) SEDs for
each star formation history.  The corresponding model parameters are
listed in Table~\ref{model}.  All the model fits consistently give a
redshift of 6.60--6.65. If we model the SED including Ly$\alpha$
emission line with an observed equivalent width of 100--300
\AA, the photometric redshift could increase up to $z \simeq 6.8$.
(This addition of a Ly$\alpha$ line changes redshift, but does not
affect the physical properties of the stellar population
significantly.)  The existence of such a Ly$\alpha$ line is still
consistent with the spectra presented by \citet{Kneib04a} since there
are a number of strong atmospheric OH lines in their wavelength range.
A redshift above 6.8, however, is unlikely because it would require
strong Ly$\alpha$ emission, which would have been seen in the
near-infrared spectrum in \citet{Kneib04a}.  This suggests that the
continuum break at 9800 \AA\ seen by \citet{Kneib04a} is either
spurious or an absorption line feature.

Figure~\ref{sed} also illustrates the importance of having {\em
Spitzer}/IRAC photometry for such a high redshift galaxy.  IRAC
measures the SED above 4000 \AA\ in the restframe, thereby
constraining the strength of the Balmer break.  The IRAC measurements
are indicative of a significant Balmer break (Figure~\ref{sed}),
suggesting that the galaxy is observed well after the most vigorous
stage of star formation (i.e., poststarburst), which is consistent
with the fact that all the exponentially decaying star formation rate
(SFR) models consistently give a galaxy age larger than the SFR
e-folding time ($\tau$).  A precise determination of the galaxy age,
however, is difficult because it essentially depends on the star
formation history we assume (Table~\ref{model}).  A reasonable lower
limit on the age seems to be $\sim 50-60$ Myr based on the $\tau=10$
Myr model (ignoring the instantaneous model, which is informative but
unrealistic).  The upper limit, on the other hand, is harder to
constrain, and the 1 Gyr burst model is consistent with an age range
of 300-700 Myrs (68\% confidence interval), the large uncertainty
being due to the slow SED evolution of this model in this age range.
Taken as a whole, the model calculations in Table~\ref{model} suggest
the existence of a mature stellar population whose age is at least 50
Myr and quite possibly a few hundred Myr.

An interesting characteristic inferred from the model fits is an
extremely blue restframe UV spectrum.  The spectral slope index
$\beta$ (defined as $f_{\lambda} \propto \lambda^{\beta}$) is $\la
-2$, as is the case with the lower redshift UDF $i$-dropout galaxies
for which NICMOS photometry is available \citep{Stanway04}.  Such a UV
spectrum is consistent with a recent starburst and a normal IMF, but
requires low extinction and/or low metallicity.  Unfortunately,
extinction and metallicity are degenerate, and are difficult to
constrain independently with our photometric resolution.  As an
example, we show this degeneracy for the case of the $\tau=100$ Myr
model in Figure~\ref{sed}b with the corresponding model parameters in
Table~\ref{model}.

By including the full range of models that produce acceptable fits in
addition to those listed in Table~\ref{model}, the various galaxy
parameters are constrained to the following ranges (instantaneous
models excluded): redshift: $6.6-6.8$; age: 50--450 Myr; stellar mass:
$0.5-1 \times 10^{9} M_{\sun}$; and SFR: 0.1--5 $M_{\sun}$ yr$^{-1}$.
These numbers reflect the ranges of the best-fit parameters without
including the parameter uncertainties associated with each fit.  The
mass is better constrained than the age, and the estimated mass ($\sim
10^{9} M_{\sun}$) is an order of magnitude smaller than those of
typical Lyman break galaxies at $z=3-4$ ($\sim 10^{10} M_{\sun}$)
\citep{Papovich01,Shapley01,Barmby04}.  However, the specific star
formation rate (SFR/$M_{star}$) \citep[e.g.,][]{Brinchmann04} is
similarly large ($> 10^{9}$ yr$^{-1}$), indicating equally vigorous
star-forming activity.

The possible detection of a mature stellar population in this galaxy
allows us to extend the argument made by \citet{Kneib04a} on the basis
of the high surface density on the sky of sources similar to that
studied here.  As implied by its discovery in a small survey area,
unless we have been extraordinarily fortunate, such sources are likely
to have a mean surface density of $\simeq$ 1 arcmin$^{-2}$ and remain
luminous over an extended period of cosmic history.  Accordingly, it is
likely that they contribute significantly to the UV photon budget
necessary for cosmic reionization.

\section{Conclusions}

The new {\em Spitzer} and {\em HST} data indicate that the redshift of
this lensed galaxy is indeed very high, probably at $z \sim 6.6-6.8$.
Comparison with a variety of stellar population synthesis models
indicates that the galaxy is in the poststarburst stage with an age of
at least $\sim$50 Myr and quite possibly a few hundred Myr, which
suggests the possibility that a mature stellar population is already
in place at such a high redshift.  Unless we have been fortunate in
its discovery, it is likely that such sources with extended lifetimes
contribute significantly to cosmic reionization.

\acknowledgments

We acknowledge helpful discussions with G.~Smith, D.~Stern, R.~Pello,
D.~Schaerer, and thank G. Neugebauer for commenting on the manuscript.
JPK acknowledges support from Caltech and CNRS.  The study of Abell
2218 as a cosmic lens is supported by NASA STScI grant
HST-GO-09452.01-A.

\clearpage

\begin{deluxetable}{lcccc}
\tablecaption{Flux densities of components $a$ and $b$ \label{flux}}
\tablewidth{0pt} 
\tablehead{ 
\colhead{Filter} & \colhead{$\lambda_{c}$\tablenotemark{a}} & 
\colhead{$\Delta\lambda$ (20\%)\tablenotemark{a}} &
\multicolumn{2}{c}{Flux density} \\
\cline{4-5} 
\colhead{} & \colhead{} & \colhead{} & \colhead{a}        & \colhead{b} \\
\colhead{} & \colhead{($\mu$m)} & \colhead{($\mu$m)} & 
\colhead{($\mu$Jy)} & \colhead{($\mu$Jy)}
} 
\startdata
$F814W$     & 0.801 & 0.242 & 0.09$\pm$0.02 & 0.03$\pm$0.01 \\
$F850LP$    & 0.915 & 0.214 & 0.39$\pm$0.02 & 0.30$\pm$0.02 \\
$F110W$     & 1.121 & 0.604 & 0.78$\pm$0.05 & 0.67$\pm$0.04 \\
$F160W$     & 1.612 & 0.418 & 0.87$\pm$0.04 & 0.80$\pm$0.04 \\
3.6 $\mu$m  & 3.561 & 0.773 & \nodata       &  1.2$\pm$0.3 \\
4.5 $\mu$m  & 4.510 & 1.015 & \nodata       &  1.0$\pm$0.2 \\ 
\enddata

\tablenotetext{a}{The effective wavelength and the full width at 20\%
of the peak calculated from the total transmission curves.}

\tablecomments{The IRAC flux densities of component $a$ are highly
  uncertain due to the subtraction of SMM-A, and therefore are not
  listed.} 

\end{deluxetable}

\begin{deluxetable}{lcclclcc}
\tabletypesize{\scriptsize}
\rotate
\tablecolumns{8}
\tablewidth{0pt}
\tablecaption{Model parameters \label{model}}
\tablehead{
\colhead{Model} & \colhead{Metallicity} & \colhead{$z$} &
\colhead{Age\tablenotemark{a}} & \colhead{Mass} & \colhead{$\tau_{V}$} & \colhead{SFR} &
\colhead{$z_{form}$} \\
\colhead{}      & \colhead{(Z$_{\sun}$)} & \colhead{} &
\colhead{(Myr)} & \colhead{(10$^{8}$ M$_{\sun}$)} &  \colhead{} &
\colhead{(M$_{\sun}$/yr)} & \colhead{}
}
\startdata
\multicolumn{8}{c}{Minimum $\chi^{2}$ models} \\ \hline
Instantaneous burst                & 0.4   & 6.60 & \phn35  (28-40)   & 5.5 & 0.25 & \nodata & 6.8 \\
e-decaying burst ($\tau = 10$ Myr) & 0.02  & 6.60 & \phn64  (56-72)   & 8.2 & 0.50 & 0.2     & 7.0 \\
e-decaying burst ($\tau = 50$ Myr) & 0.4   & 6.60 & 102  (82-114)  & 6.3 & 0.25 & 2.2     & 7.3 \\
e-decaying burst ($\tau= 100$ Myr) & 1.0   & 6.65 & 181  (150-201) & 8.3 & 0.0  & 2.0     & 8.1 \\
1 Gyr burst                        & 1.0   & 6.65 & 453  (287-693) & 9.7 & 0.0  & 2.6     & 12.3 \\
\cutinhead{e-decaying ($\tau = 100$ Myr) models with sub-solar metallicities}
e-decaying burst ($\tau = 100$ Myr)& 0.005 & 6.60 & 161  (121-185) & 11  & 0.75 & 3.3 & 7.8 \\
                                   & 0.02  & 6.60 & 143  (124-175) & 8.5 & 0.50 & 3.2 & 7.7 \\
                                   & 0.2   & 6.60 & \phn90 (66-120)  & 5.4 & 0.50 & 4.2 & 7.2 \\
                                   & 0.4   & 6.60 & 128  (105-156) & 6.2 & 0.25 & 2.8 & 7.5 
\enddata 

\tablenotetext{a}{The numbers in parentheses indicate the 68\% confidence
  interval based on $\Delta\chi^{2}$.}
\tablecomments{Models with GALAXEV \citep{Bruzual03} explored the
  following range of parameters: (1) Initial mass function (IMF):
  Salpeter with the lower and upper mass cutoffs of 0.1 and 100
  M$_{\sun}$; (2) metallicity: 0.005, 0.02, 0.2, 0.4, 1.0, and 2.5
  Z$_{\sun}$; (3) dust extinction: $\tau_{V}=$ 0.0, 0.25, 0.5, 0.75,
  1.0, 1.25, and 1.5, where $\tau_{V}$ is the total effective $V$-band
  optical depth seen by young (age $< 10^{7}$ yrs) stars.  The optical
  depth toward old stars (age $ > 10^{7}$ yrs) was set to be 1/3 of
  this. See \citet{Charlot00} and \citet{Bruzual03} for more detail;
  and (4) star formation histories: instantaneous burst, exponentially
  decaying burst (SFR $\propto e^{-t/\tau}$, where $t$ is the galaxy
  age and $\tau$ is the SFR e-folding time scale) with $\tau =$ 10,
  50, and 100 Myr, and 1 Gyr burst (a constant SFR model with the SFR
  scaled such that all the gas gets consumed in 1 Gyr).
  The flux below Ly$\alpha$ ($\lambda_{rest}=1216$ \AA) was set to
  zero, and no Ly$\alpha$ emission line was included.  SFRs listed are
  intrinsic model values.  We assumed a magnification factor of 25
  \citep{Kneib04a} and the cosmological parameters of $\Omega_{\rm
  M}=0.3$, $\Omega_\Lambda=0.7$ and H$_0$=70 km~s$^{-1}$~Mpc$^{-1}$. }

\end{deluxetable}

\end{document}